\documentclass[conference]{IEEEtran}

\usepackage[dvips]{color}
\usepackage{epsf}
\usepackage{times}
\usepackage{epsfig}
\usepackage{graphicx}
\usepackage{float} 
\usepackage{amsmath}
\usepackage{amssymb}
\usepackage{amsxtra}
\usepackage{amsthm}
\usepackage{bbm}
\usepackage{xcolor}
\usepackage{soul}

\usepackage{here}
\usepackage{rawfonts}
\usepackage{times}
\usepackage{url}
\usepackage{cite}
\usepackage{comment}
\usepackage[utf8]{inputenc}
\usepackage{caption}
\usepackage{subcaption}
\usepackage[normalem]{ulem}

\usepackage{pstricks}
\usepackage{algorithm}
\usepackage{algpseudocode}
\usepackage{lipsum}

\IEEEoverridecommandlockouts 

\begin{document}

\title{\huge
	Domain-Generalizable Semantic Communications With Pilot-Assisted  Adaptation over  Wireless Channels
}\vspace{-0.2cm}	
\author{ 
	\IEEEauthorblockN{ 
	Haoran Chang$^\ast$, Sweekar Piya$^+$, Xiajun Jiang$^+$, Qianqian Zhang$^\ast$} 
	
	\IEEEauthorblockA{\small
		$^\ast$Department of Electrical and Computer Engineering, Rowan University,  NJ, USA,
		Emails: \url{{changh35,zhangqia}}@rowan.edu \\
        $^+$Department of Computer Science, Rowan University,  NJ, USA,
		Emails: \url{{piyasw25,jiangx}}@rowan.edu 
        \vspace{-0.2cm}				
	}
}
\maketitle

\begin{abstract} 
Semantic communication offers a promising approach to support intelligent services in next-generation wireless networks by efficiently transmitting task-relevant information. 
However, its performance can degrade significantly when input data and wireless channel conditions differ from those encountered during training. 
In this paper, we propose a robust semantic communication framework that addresses these distributional  mismatches through two independent but complementary modules. 
At the transmitter, a domain-generalization module extracts task-relevant features while suppressing domain-specific information.
At the receiver, a pilot-assisted channel characterization module extracts channel context from received pilot signals to guide semantic recovery under varying wireless conditions. 
Through the end-to-end training of semantic encoder-decoder pair with domain alignment and channel adaptation models, the proposed framework improves task performance under simultaneous input and channel distribution shifts with limited transmitter-receiver coordination during deployment. 
Simulation results show that the proposed framework improves classification accuracy by more than $18\%$ over a conventional semantic communication baseline in the unseen  scenarios.

\end{abstract}  
 

\section{Introduction}
 
Semantic communication has emerged as a promising foundation for next-generation wireless networks. 
By extracting and transmitting task-relevant information from the communication input data, semantic communication enables efficient task execution at the receiver while reducing communication overhead \cite{shao2022task}.  
Recent advances in deep learning have enabled the joint design of semantic feature extraction and wireless transmission, allowing learned representations to preserve information essential for downstream tasks under limited communication resources. 
These capabilities offer substantial potential for applications such as autonomous driving, remote monitoring, and intelligent manufacturing, where timely and reliable inference is critical. 

However, the effectiveness of learning-based semantic communication depends heavily on the conditions under which the system is trained \cite{SCDA_1}. 
Models trained offline on specific datasets and channel distributions may experience substantial performance degradation when deployed in real-world environments. 
In particular, variations in communication input data can compromise semantic feature extraction, while changes in wireless propagation can distort the transmitted representations in ways not captured during training. 
These mismatches in the communication input and wireless channel distributions affect different stages of the communication process, limiting the generalization and reliability of the overall system. 
Addressing both forms of distribution shift is therefore essential for translating the potential of semantic communication into reliable task performance in practical wireless environments.

Existing studies have explored adaptive frameworks to improve the robustness of semantic communication. 
To mitigate channel impairments and support robust semantic decoding, \cite{kim2024pilot} applied the masked autoencoders for reliable channel estimation with reinforcement learning for pilot allocation.
Similarly, \cite{yang2021ofdm} integrated  OFDM processing, channel estimation, and equalization into deep joint source–channel coding to improve image transmission over multipath fading channels. 
Beyond channel estimation and compensation, \cite{xu2022attention} introduced attention modules to adapt semantic transmission over wireless channels with varying signal-to-noise ratios (SNRs). 
Although these approaches enhance robustness to channel variations, they do not explicitly address input distribution shifts that can compromise semantic feature extraction.  
To this end, \cite{zhang2023dynamic} employed an adaptation network to map incoming images back to the original data domain in which the semantic communication system is pretrained, and \cite{SCDA_1} extended 
this approach to support adaptation across multiple domains. 
However, both works only focus on domain-specific mappings, which only improves semantic transmission under domain shifts encountered during training 
but offers limited robustness to unseen distributions or scenarios.   
Furthermore, \cite{nguyen2025crossdomain} used a federated learning framework to combine global representation alignment and thus mitigate data heterogeneity among multiple clients. 
Recent work in \cite{mao2026sfda}  proposed a knowledge-base-enabled adaptation framework to improve joint image reconstruction and task performance under domain shifts and varying link quality. 
Despite these advances, existing approaches often rely on  channel feedback, domain-specific adaptation, or iterative  transmitter-receiver  coordination, introducing additional overhead that can limit deployment efficiency and practical applicability.  

In this paper, we propose a robust semantic communication framework that jointly addresses input data shifts and wireless channel variations. 
The framework comprises two independent yet complementary modules for semantic feature extraction at the transmitter and channel compensation at the receiver.  
First, a domain-generalization module is introduced to enable the semantic transmitter to detect input distribution shifts and suppress domain-specific information. 
This module generates domain-invariant semantic representations while preserving task-relevant information.  
Second, a pilot-assisted channel characterization module is developed to adapt the semantic receiver to varying wireless environments. 
This module extracts channel context from  received pilot signals and generates a conditioning vector to guides the recovery of task-relevant information from received semantic signals. 
Together, these modules  improves task performance of the semantic communication system under simultaneous input and channel distribution shifts, with limited transmitter-receiver coordination during deployment. 
Simulation results show that the proposed framework improves classification accuracy by more than $18\%$ compared with the traditional semantic communication framework in the unseen communication scenarios.  



The rest of this paper is organized as follows. 
Section II introduces the system model and problem formulation. 
Section III presents the proposed  solution for adaptive semantic communications. 
Simulation results are shown in Section IV, and conclusions are drawn in Section V.

\section{System Model and Problem Formulation}\label{sysModel_proFormulation}

\subsection{System Model}  

\begin{figure*}[t]  
    \centering
    \makebox[\textwidth][l]{ 
        \includegraphics[width=1\textwidth]{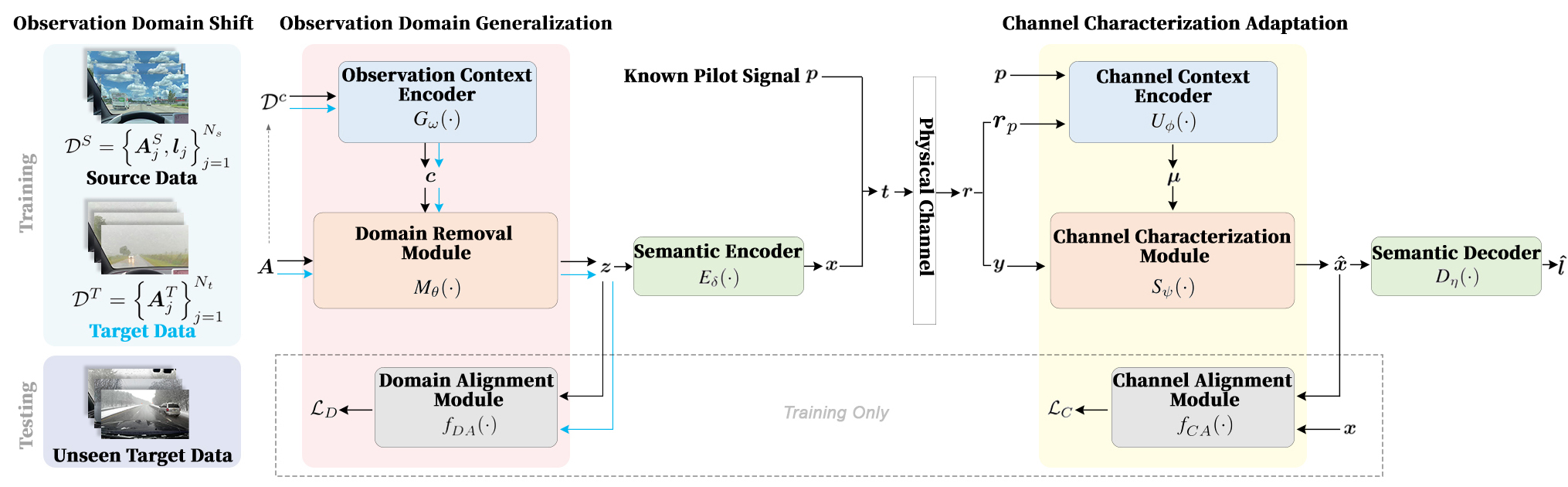}
    }
    \vspace{-0.4cm}
    \caption{System architecture of domain-generalizable semantic communications with wireless channel adaptation.}
    \vspace{-0.2cm}
    \label{fig:system_model} 
\end{figure*}





We consider a  semantic communication system in which a transmitter extracts task-relevant semantic information from input images and conveys it over a wireless channel to a receiver for task execution. 
For example,   a self‑driving vehicle captures images of its surroundings using an onboard camera and transmits the extracted semantic information to a roadside unit, as illustrated in Fig. \ref{fig:system_model}. 
Typically, such a task-oriented  semantic communication system is trained offline, but real-world deployment can introduce two distinct types of distribution shifts relative to the training conditions.    
First, variations in weather, illumination, and camera configurations alter the distribution of input images. We refer to this as \emph{observation domain shift}. 
Second,  changes in the propagation environment and device mobility alter the wireless channel distribution, such as multipath propagation, fading, and Doppler shifts.  We refer to this as \emph{channel characterization shift}. 
These shifts together affect both semantic feature extraction and wireless transmission. 
Without effective adaptation mechanisms, distribution shifts can significantly degrade the task performance when the deployed system encounters visual or channel conditions not represented in training.  
To address this challenge, we introduce a generalizable framework  with four components, including an observation context encoder  and an domain removal module  at the transmitter as defined in Section \ref{semantic_tx}, and a channel context encoder and a characterization module at the receiver as defined in Section \ref{semantic_rx}.

\subsubsection{Cross-domain Dataset}
To generalize the encoder's capacity to the unseen image input, two training datasets $\mathcal{D}^S$ and $\mathcal{D}^T$ are used.  
In particular,  $\mathcal{D}^S=\{  \boldsymbol{A}^{S}_j, {l}_j \}_{j=1,\cdots,N_s} $ is a \emph{source-domain} dataset that contains a large amount of training images from a lab setting or public dataset with labels ${l}_j \in \{1 \ldots,  L\}$, where $L$ denotes the total number of task classes.  
Meanwhile, $\mathcal{D}^T = \{  \boldsymbol{A}^{T}_j\}_{j=1, \cdots, N_t}$    is a \emph{target-domain} dataset with limited and unlabeled training images from the real-world application scenarios. 
In particular, we consider $I$ different target domains, 
and the training dataset can be given as $\mathcal{D}=\mathcal{D}^S \cup \{\mathcal{D}^{T}_i \}_{i=1,\cdots,I}$. 
Any two domains is required to keep a minimum distribution discrepancy, measured by the Kullback–Leibler (KL) divergence, as $\ D_{\mathrm{KL}}(\mathcal{D}_i || \mathcal{D}_v) \ge \epsilon$ for all $\mathcal{D}_i \neq \mathcal{D}_v \in \{\mathcal{D}^S, \mathcal{D}^{T}_1, \ldots, \mathcal{D}^{T}_I\}$.   
To enable the semantic transmitter to detect and handle distribution shifts, a \emph{context} dataset $\mathcal{D}^c$ will be sampled from $\mathcal{D}$ conditioned on the domain of the new image input. 
Finally, in the testing stage, an unseen target-domain dataset with unknown image distributions will be used to evaluate the performance of the generalizable semantic communication system, as presented in  Section \ref{Simulation}. 
For example,  a source-domain dataset consists exclusively of clear-weather images annotated for vehicle and pedestrian detection, whereas  target-domain images are captured under adverse weather conditions, such as heavy rain or snow, where data collection and annotation are highly costly.

\subsubsection{Domain-generalizable Semantic Transmitter} \label{semantic_tx}

In the proposed framework, an image input $\boldsymbol{A} \in \mathbb{R}^{C\times H \times W}$ will be first processed by the domain removal module to mitigate distribution shifts and output a domain-invariant representation $\boldsymbol{z} \in \mathbb{R}^{d_{z}}$   before semantic encoding. 

To facilitate domain-signature removal,    a context dataset $\mathcal{D}^c$ is constructed by sampling a small number of representative images from $\mathcal{D}$ conditioned on the domain of the current input $\boldsymbol{A}$. 
The observation context encoder $G_{\omega}(\cdot)$ extracts auxiliary domain information from $\mathcal{D}^c$ as 
\begin{equation}
    \boldsymbol{c} = \textit{G}_{\omega}(\mathcal{D}^c|\boldsymbol{A}),
\end{equation}
where $\omega$ is the trainable parameter and $\boldsymbol{c} \in \mathbb{R}^{d_{c}}$ characterizes the observation domain of $\boldsymbol{A}$. 
Next, the input image $\boldsymbol{A}$ is processed by the domain removal module $\textit{M}_{\theta}(\cdot)$ as 
\begin{equation}
    \boldsymbol{z} = \textit{M}_{\theta}(\boldsymbol{A}|\boldsymbol{c}), 
\end{equation}
where $\theta$ is the   trainable parameter  and $\boldsymbol{z} \in \mathbb{R}^{d_{z}}$ is the domain-invariant feature  that preserves task-relevant information. 

During offline training, the  representation  $\boldsymbol{z}$ will be fed into the domain alignment module $\textit{f}_{DA}(\cdot)$ to quantify the residual discrepancy between the source and target domains. 
The domain alignment loss is defined as 
\begin{equation} \label{equ_LD_1}
\begin{aligned}
    &\mathcal{L}_{D} (\omega,\theta)  =  \\
    &f_{DA} \left( {M}_{\theta}\left(\boldsymbol{A}^{S} \bigm| G_{\omega}(\mathcal{D}^c|\boldsymbol{A}^{S}) \right),
    M_{\theta} 
    \left(\boldsymbol{A}^{T}\bigm|G_{\omega}(\mathcal{D}^c|\boldsymbol{A}^{T}) \right) \right),
\end{aligned}
\end{equation}
where $f_{DA}$ measures the discrepancy between the source and target distributions.  
Minimizing $\mathcal{L}_{D}$ encourages $M_{\theta}(\cdot)$ to produce domain-invariant features, as detailed in Section \ref{solution_oda}. 
After training, $\boldsymbol{z}$ is passed to semantic encoder $E_{\delta}(\cdot)$ to be mapped into a vector of complex-valued symbols $\boldsymbol{x} \in \mathbb{C}^{k}$: 
\begin{equation}
    \boldsymbol{x} = E_{\delta}(\boldsymbol{z}).
\end{equation}

\subsubsection{Channel-adaptive Semantic Receiver} \label{semantic_rx}

During deployment, the wireless  environment can differ substantially from the offline training setting. 
Consequently, transmission over an unknown wireless channel can induce a shift in the received signal distribution relative to the end-to-end training conditions. 
To this end, the channel context encoder is introduced at the receiver to extract channel characteristics from received pilot signals, and then the channel characterization module uses the extracted information to suppress channel-specific features in the received signal, so as to mitigate the effects of channel distribution shifts before semantic decoding.

Specifically, to characterize the wireless channel, we concatenate a known pilot sequence $\boldsymbol{p} \in \mathbb{C}^{k_p}$  with the semantic-encoded symbols $\boldsymbol{x}$ to form the transmit vector $\boldsymbol{t} = [\boldsymbol{p}^\top, \boldsymbol{x}^\top]^\top\in \mathbb{C}^{k_p+k}$. 
Assuming a block flat-fading channel that remains constant over the pilot and data transmission, the received signal is 
\begin{equation}
    \boldsymbol{r} = \boldsymbol{h} \odot \boldsymbol{t} + \boldsymbol{n},
\label{recieve}
\end{equation}
where $\boldsymbol{h}$ denotes the flat-fading channel, $\odot$ denotes element-wise multiplication, and $\boldsymbol{n} \sim \mathcal{CN}(0, \sigma_n^{2} \mathbf{I})$ is additive white Gaussian noise (AWGN).  
The received vector is partitioned as $\boldsymbol{r}=[\boldsymbol{r}_p^\top,\boldsymbol{y}^\top]^{\top}$, where $\boldsymbol{r}_p\in\mathbb{C}^{k_p}$ and $\boldsymbol{y}\in\mathbb{C}^{k}$ contain the received pilot and data symbols, respectively. The received pilot sequence $\boldsymbol{r}_p$, together with the known sequence $\boldsymbol{p}$, is then fed into the channel context encoder $U_{\phi}(\cdot)$ as 
\begin{equation}
    \boldsymbol{\mu} = U_{\phi}(\boldsymbol{p}, \boldsymbol{r}_p), 
\end{equation}
where $\phi$ is the trainable parameter and $\boldsymbol{\mu} \in \mathbb{R}^{d_{\mu}}$ is a condition vector that captures channel-specific information. 
Subsequently, $\boldsymbol{\mu}$ and $\boldsymbol{y} \in \mathbb{C}^{k}$ are jointly processed by the channel characterization module $\textit{S}_{\psi}(\cdot)$ 
to recover the channel-invariant $\boldsymbol{\hat{x}} \in \mathbb{C}^{k}$, via:
\begin{equation}
    \boldsymbol{\hat{x}} = \textit{S}_{\psi}(\boldsymbol{y}, \boldsymbol{\mu}).
\end{equation}

During training, the transmitted symbols $\boldsymbol{x}$ and the channel-compensated representation $\hat{\boldsymbol{x}}$ 
are  fed into the channel alignment module $\textit{f}_{CA}(\cdot)$ to compute the channel alignment loss: 
\begin{equation} \label{equ_channel_loss}
    \mathcal{L}_{C}(\psi, \phi) = \textit{f}_{CA}\Bigl( \boldsymbol{x}, \textit{S}_{\psi}(\boldsymbol{y},\textit{U}_{\phi}(\boldsymbol{p},\boldsymbol{{r}_p}))\Bigr).
\end{equation}
Minimizing $\mathcal{L}_{C}$ encourages $S_{\psi}(\cdot)$ to suppress channel-dependent distortions and recover the transmitted semantic symbols, as detailed in Section \ref{solution_cca}. 
After training, $\boldsymbol{\hat{x}}$ is passed to the semantic decoder $D_{\eta}(\cdot)$, with   trainable parameter $\eta$, to predict the task label as  
\begin{equation}
    \hat{l} = D_{\eta}(\boldsymbol{\hat{x}}). 
\end{equation}

For task-oriented semantic communication, the end-to-end training objective is to minimize  the cross-entropy loss \cite{crossentropy} for task-label prediction on the source-domain dataset: 
\begin{equation}
    \mathcal{L}_{task} (\omega, \theta, \delta, \phi, \psi,\eta) =  
        \textit{f}_{CE}( l , \hat{l}).  
\label{L_task}
\end{equation}

To achieve accurate semantic task execution across diverse observation domains and channel conditions, the overall loss function is defined as
\begin{equation}
    \mathcal{L}_{total} (\omega, \theta, \delta, \phi, \psi,\eta) =  
    \mathcal{L}_{task} + \lambda\mathcal{L}_{D} +\gamma\mathcal{L}_{C} 
    , 
\label{L_total}
\end{equation}
where $\lambda$ and $\gamma>0$ are hyperparameters to balance the learning objective between task performance with domain alignment and channel characterization.  
Consequently, the optimization problem can be formulated as
\begin{subequations} 
    \begin{align} 
        \underset{ \substack{\omega, \theta, \delta,  \phi, \psi,\eta } }   {\min} &
        \mathbb{E}_{\boldsymbol{A}\sim \mathcal{D}}  \left[ \mathcal{L}_{total} (\omega, \theta, \delta, \phi, \psi,\eta)   \right]   \\ 
        \text{s.t.}~ &  \frac{\mathbb{E} [ ||\boldsymbol{h} \odot \boldsymbol{t} (\omega, \theta, \delta)||^2 ]} { \sigma_n^2} \ge \tau_{\text{thre}},  \label{con_snr}  \\
                     &  {||\boldsymbol{t} (\omega, \theta, \delta)||^2} \le P_{\text{max}}.  \label{con_tp} 
    \end{align}\label{equ_pro}
\end{subequations}  
Constraint (\ref{con_snr}) requires SNR to be no less than  $\tau_{\text{thre}}$ and (\ref{con_tp}) limits the transmit power to $P_{\text{max}}$. 
Problem (\ref{equ_pro}) is challenging to solve for three reasons. 
First, the information bottleneck imposed by the wireless channel limits semantic communication performance, necessitating joint encoder-decoder optimization under transmission constraints. 
In particular, under limited power budget  and poor channel SNR, 
the semantic encoder must effectively extract task information, while the semantic decoder needs to accurately recover semantic message for reliable task execution. 
Second, observation-domain shifts alter the input image distributions, hindering robust semantic feature extraction and degrading task performance. 
Third, dramatic changes in wireless channel distributions  introduce unfamiliar signal distortions, compromising the semantic receiver’s reliability. 
Addressing these challenges requires jointly learning on domain-invariant semantic encoding and channel-adaptive decoding to sustain task performance across diverse observation domains and channel conditions. 

 
\section{Proposed Solution} \label{solution}

This section details the proposed semantic communication framework, including the observation domain generalization and the channel characterization adaptation, with the objective to improve  task performance under the real-world deployment. 

\subsection{Observation Domain Generalization} \label{solution_oda}

To handle the evolving distribution in the observation domain, the proposed  framework is designed upon the observation context encoder and the domain removal module, to achieve dynamic domain adaptability.  
First, observation context encoder $\textit{G}_{\omega}$ generates domain knowledge  $\boldsymbol{c}$ based on the input.   
Next, domain removal module $\textit{M}_{\theta}$ uses  $\boldsymbol{c}$ as condition to generate a domain-invariant semantic representation $\boldsymbol{z}$.  

\subsubsection{Observation context encoder}
This encoder is implemented as a convolutional neural network (CNN) that processes a small set of context images $\mathcal{D}^c$ sampled from $\mathcal{D}$ under the same domain category as the current input $\boldsymbol{A}$.  
The context images are processed as a batch, and their extracted domain features  are averaged via mean pooling to obtain a unified domain representation $\boldsymbol{c}$.






\subsubsection{Domain removal module} 
This module comprises $B$ layers, each containing a CNN-based feature extraction network with trainable parameters $\theta_b^F$ and a feature-wise linear modulation (FiLM)\cite{film} generator with trainable parameters $\theta_b^R$, for $b=1,\ldots,B$. 
The  complete parameter set  is thus denoted by $\theta=\{\theta_b^F,\theta_b^R\}_{\forall b}$. 
Fig. \ref{fig:DRM} illustrates the architecture of domain removal module, with the first layer clearly shown and subsequent  layers  following the same structure. 
 
At each layer $b$ of the domain removal module, a layer-specific FiLM generator $\textit{Q}_{\theta^F_b}(\cdot)$, implemented as a multilayer perceptron (MLP),  maps the domain representation $\boldsymbol{c}$ into two modulation vectors: $(\boldsymbol{\alpha}_b,\boldsymbol{\beta}_b)= \textit{Q}_{\theta^R_b}(\boldsymbol{c})$, 
where $\boldsymbol{\alpha}_b, \boldsymbol{\beta}_b\in\mathbb{R}^{C_{b}}$ denote  feature-wise scaling and shifting, respectively. 
In parallel, as shown in Fig. \ref{fig:DRM}, the first-layer feature extraction network of the domain removal module, parameterized by $\theta_1^R$, maps the input image $\boldsymbol{A}$ to an intermediate feature map 
$\boldsymbol{F}_1  \in \mathbb{R}^{C_1 \times H_1 \times W_1}$.  
FiLM   then further process $\boldsymbol{F}_1$ as 
\begin{equation}
    \widetilde{\boldsymbol{F}}_1 = \boldsymbol{\alpha}_1 \odot\boldsymbol{F}_1 + \boldsymbol{\beta}_1.  
\end{equation}
Here, $\widetilde{\boldsymbol{F}}_1$ preserves the same dimensionality as the intermediate feature map $\boldsymbol{F}_1$. 
Subsequent layers, indexed by $b=2,\ldots,B$,  follow the same procedure, using layer-specific parameter $\theta^R_b$ and taking $\widetilde{\boldsymbol{F}}_{b-1}$ as input.  
Note that, our design allows FiLM modulation to be bypassed at any layer $b$ by setting  $\boldsymbol{\alpha}_b=\boldsymbol{1}$ and $\boldsymbol{\beta}_b=\boldsymbol{0}$. 
Finally, the output of the  domain removal module is given as $\boldsymbol{z} = M_{\theta}(\boldsymbol{A}|\boldsymbol{c})$. 



\begin{figure}[t]  
    \centering
    \includegraphics[width=0.47\textwidth]{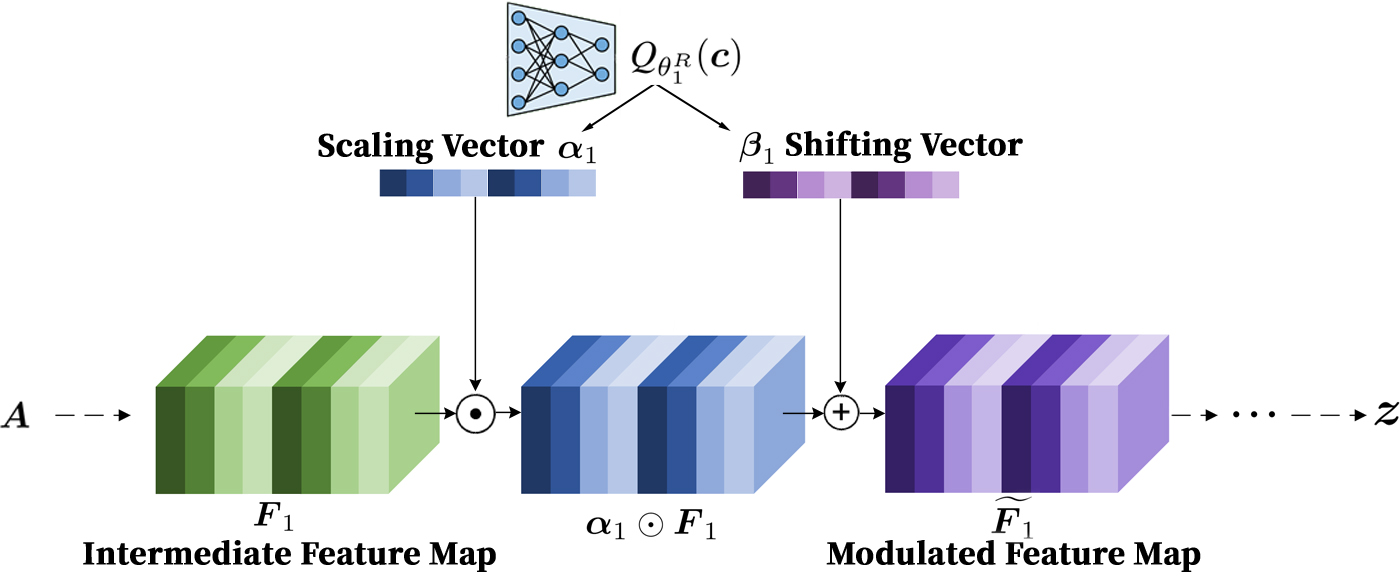} 
    \caption{\label{fig:DRM}The context-conditioned FiLM layer within the CNN backbone of the domain removal module.} 
    \label{fig:DRM}
    \vspace{-0 cm}
\end{figure}



\subsubsection{Training objective of domain alignment} 
To encourage  domain-invariant  in $\boldsymbol{z}$,  we introduce a CORAL-based domain alignment module \cite{coral} that minimizes source–target discrepancy by aligning feature covariance matrices. 
Specifically, let $\boldsymbol{Z}_S = [ \textit{M}_{\theta}(\boldsymbol{A}_{1}^{S}),\, \ldots,\, \textit{M}_{\theta}(\boldsymbol{A}_{N_s}^{S}) ]$ and $\boldsymbol{Z}_T = [ \textit{M}_{\theta}(\boldsymbol{A}_{1}^{T}),\, \ldots,\, \textit{M}_{\theta}(\boldsymbol{A}_{N_t}^{T}) ]$ 
denote the source- and target-domain features produced by domain removal module, respectively.  
The domain alignment loss ${f}_{DA}(\cdot)$  measures the discrepancy between their covariance matrices as 
\begin{equation}
    \boldsymbol{O}_S =
    \frac{1}{N_s-1}
    \left(
    \boldsymbol{Z}_S^\top\boldsymbol{Z}_S
    -
    \frac{1}{N_s}
    (\boldsymbol{1}^\top\boldsymbol{Z}_S)^\top
    (\boldsymbol{1}^\top\boldsymbol{Z}_S)
    \right),
\end{equation}
\begin{equation}
    \boldsymbol{O}_T =
    \frac{1}{N_t-1}
    \left(
    \boldsymbol{Z}_T^\top\boldsymbol{Z}_T
    -
    \frac{1}{N_t}
    (\boldsymbol{1}^\top\boldsymbol{Z}_T)^\top
    (\boldsymbol{1}^\top\boldsymbol{Z}_T)
    \right). 
\end{equation}
Accordingly, the domain loss $\mathcal{L}_{D}$ in (\ref{equ_LD_1}) can be expressed  as
\begin{equation}
    \mathcal{L}_{D}(\omega, \theta)
    =
    \frac{1}{4d_z^2}
    \left\|
    \boldsymbol{O}_S-\boldsymbol{O}_T
    \right\|_F^2,
\label{L_D}
\end{equation}
where $\left\| \cdot \right\|_F^2$ represents the squared matrix Frobenius norm.


\setlength{\textfloatsep}{6pt}
\begin{algorithm}[t]
\caption{\label{alg1}Train the Proposed Framework}
\begin{algorithmic}[1]\small

\State \textbf{Input}:  
$\mathcal{D}$, 
$\boldsymbol{p}$, 
$\lambda$, $\gamma$, 
$\tau_{\text{thre}}$, and $P_{\text{max}}$ 

\State \textbf{Initialization}:
Initialize $\omega$, $\theta$, $\delta$, $\phi$, $\psi$, and $\eta$

\While{$\mathcal{L}_{total}$ in (\ref{L_total}) has not converged}

        \State Sample $N$ mini-batches from $\mathcal{D}$

        \For{$n=1,\ldots,N$}
                \State Sample $(\boldsymbol{A}^{S},{l})$ 
                and $\boldsymbol{A}^{T}$ from mini-batch $n$
    
                \State \textbf{Observation Domain Generalization}:
                
                \State  \hspace{0.2cm}  $\boldsymbol{z}^{S}
                \gets {M}_{\theta}
                (\boldsymbol{A}^{S} | G_{\omega}( \boldsymbol{A}^{S}))$,  $\boldsymbol{z}^{T}
                \gets {M}_{\theta}
                (\boldsymbol{A}^{T} | G_{\omega}(\boldsymbol{A}^{T}))$ 
                \State \hspace{0.2cm} Compute and update $\mathcal{L}_D$ by (\ref{L_D}) 

                \State \textbf{Semantic Encoder}:
                \State \hspace{0.2cm} $\boldsymbol{x}
                \gets E_{\delta}(\boldsymbol{z}^{S})$
                and $\boldsymbol{t}
                \gets [\boldsymbol{p}^{\top},\boldsymbol{x}^{\top}]^{\top}$ 
                \State \hspace{0.2cm} Adjust transmit power based on (\ref{con_snr}) and (\ref{con_tp})
                \State \hspace{0.2cm} Transmit $\boldsymbol{t}$ over the channel

                \State \textbf{Channel Characterization Adaptation}:
                \State \hspace{0.2cm}
                Receive $\boldsymbol{r}$ via (\ref{recieve}) and $(\boldsymbol{r}_p,\boldsymbol{y})
                \gets \operatorname{Split}(\boldsymbol{r})$
                \State \hspace{0.2cm} $\hat{\boldsymbol{x}}
                \gets {S}_{\psi}
                \left(\boldsymbol{y}, U_{\phi}(\boldsymbol{p},\boldsymbol{r}_p) \right)$ 
                
                \State \hspace{0.2cm} Compute and update $\mathcal{L}_C$ by (\ref{L_C})
                
                \State \textbf{Semantic Decoder}:
                
                \State \hspace{0.2cm} $\hat{l} \gets D_{\eta}(\boldsymbol{\hat{x}})$  and compute $\mathcal{L}_{task}$ by (\ref{L_task})
        \EndFor
        
        \State \textbf{Overall Optimization}:
        \State \hspace{1em} $\mathcal{L}_{total}
        \gets \mathcal{L}_{task}
        +\lambda\mathcal{L}_{D}
        +\gamma\mathcal{L}_{C}$
        \State \hspace{1em} Update $\omega$, $\theta$, $\delta$, $\phi$, $\psi$, and $\eta$
        using gradient descent.

\EndWhile

\State \textbf{Output}: $G_{\omega}(\cdot)$,
$M_{\theta}(\cdot)$, $E_{\delta}(\cdot)$,
$U_{\phi}(\cdot)$, $S_{\psi}(\cdot)$, and
$D_{\eta}(\cdot)$

\end{algorithmic}

\end{algorithm}

\subsection{Channel Characterization Adaptation} \label{solution_cca}

To adapt to real-world wireless link variations,  the channel context encoder $U_{\phi}$ employs an MLP to extract a channel conditioning vector from the known pilot signal $\boldsymbol{p}$ and its received counterpart $\boldsymbol{r}_p$, yielding $\boldsymbol{\mu}=U_{\phi}(\boldsymbol{p},\boldsymbol{r}_p)$. 
This vector guides the channel characterization module $S_{\psi}$, also implemented as an MLP, to recover the semantic representation $\boldsymbol{\hat{x}} = \textit{S}_{\psi}(\boldsymbol{y}, \boldsymbol{\mu})$ from the received signal $\boldsymbol{y}$.  
Since channel impairments are generally not reversible, our goal is to preserve the task-relevant  information in $\hat{\boldsymbol{x}}$ rather than  reconstruct symbols exactly. 
To evaluate the accuracy of task-information recovery, the channel alignment loss in (\ref{equ_channel_loss}) can be expressed as 
\begin{equation}
    \mathcal{L}_C(\phi, \psi) =
    D_{\mathrm{KL}} ( \boldsymbol{q}_\eta(\boldsymbol{x}) || \boldsymbol{q}_\eta(\hat{\boldsymbol{x}}) ), 
\label{L_C}
\end{equation}
where $ \boldsymbol{q}_\eta(\boldsymbol{x})$ and $\boldsymbol{q}_\eta(\hat{\boldsymbol{x}}) \in \mathbb{R}^{L}$  are the predicted probability vectors over $L$ task classes 
for transmitted and recovered representations, respectively. 
Finally, the semantic decoder $D_{\eta}(\cdot)$ is jointly trained with all aforementioned modules using the loss function in (\ref{L_task}) to optimize end-to-end task performance. 
Algorithm \ref{alg1} summarizes the overall training procedure, integrating semantic encoder–decoder optimization, observation-domain generalization, and channel adaptation mechanisms.

\section{Simulation} \label{Simulation}

In our simulations, we employ three groups of data for training and evaluation: a source-domain, seen target-domains, and an unseen target domain.  
Here, CIFAR-10  \cite{cifar10} serves as the source-domain dataset, comprising $60,000$ images across $10$ object categories.  
Meanwhile, we use brightness-corrupted images from CIFAR-10-C  \cite{cifar10c} to serve as target-domain data.  
Therefore,   domain shifts are defined as different levels of brightness corruption in the input image. 
In particular, severity levels $1$–$4$ of brightness corruption are included as seen target domains, with $60\%$, $10\%$, and $30\%$ of the images  for training, validation, and testing, respectively. 
And severity level $5$ is reserved exclusively for testing as the unseen target domain. 

For the observation context encoder, we adopt ResNet-18 ~\cite{resnet18} as the CNN backbone to extract a  feature vector $\boldsymbol{c}$ as context information with $d_c=512$.  
Similarly, the domain removal module employs ResNet-18 with a FiLM layer incorporated after each of its four residual stages ($B=4$).  
These FiLM layers allow the domain context vector $\boldsymbol{c}$ to guide the generation of a  domain-invariant feature vector $\boldsymbol{z}$ with $d_z=512$. 
The semantic encoder and decoder are implemented using an autoencoder framework, with an adjustable compression rate (CR) between $0$ and $1$, and the semantic encoder generates a transmission vector of $k$ complex-valued symbols with $k = \mathrm{CR} \times d_z$. 
A pilot signal with $k_p = 512$ is employed for channel characterization. 
The entire framework is trained over AWGN channels, with SNRs from $-15$ to $6$ dB. 
The loss weighting coefficients are $\lambda = 0.1$ and $\gamma=0.01$. 

\begin{figure}[t] \vspace{-0.2cm}
    \centering
    \includegraphics[width=9.2cm]{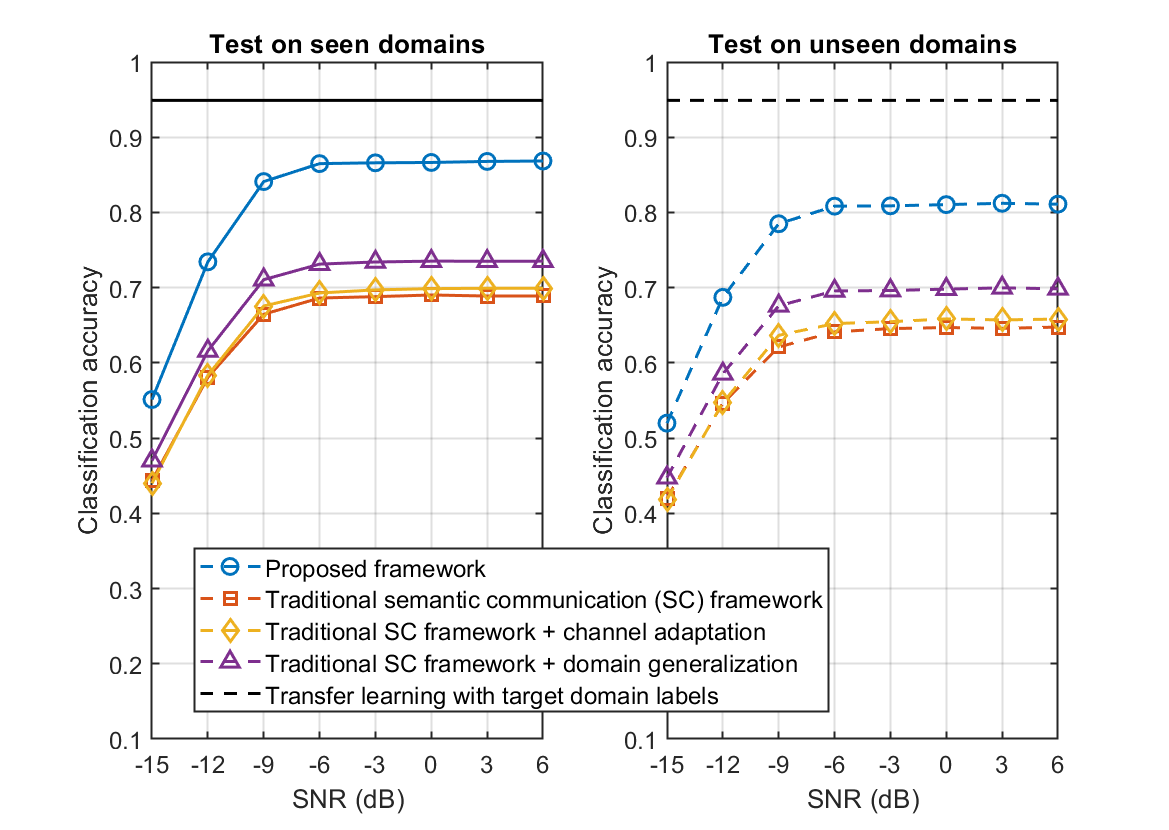}\vspace{-0.1cm}
    \caption{Testing performance on classification accuracy versus SNR for seen and unseen target domains.}
    \label{fig:accuracy}    \vspace{-0 cm}
\end{figure}

Fig. \ref{fig:accuracy} presents  image classification accuracy versus SNR for the seen and unseen target domains  at a fixed CR $=0.1$.  
The black line represents a transfer learning baseline, which requires task labels from the source domain and all target domains, including those unseen by the proposed method during training. 
Wireless transmission is omitted to provide an idealized reference for classification accuracy in the transfer-learning case.  
The red line with square markers represents the traditional semantic communication framework without observation domain generalization or channel adaptation. 
The proposed framework, represented by the blue line with circular markers, achieves more than $18\%$ higher classification accuracy than the traditional framework under both seen and unseen target domains. 
This improvement is contributed by the joint use of observation domain generalization and channel characterization adaptation, which mitigate input distribution shifts and channel impairments, respectively, thereby improving task performance across different observation domains and wireless channel conditions.
The yellow line with diamond markers represents the framework employing only channel characterization adaptation. 
Without observation domain generalization, its performance is significantly lower than the proposed framework. 
However, it still outperforms the traditional method, as channel characterization adaptation helps the semantic decoder to mitigate channel-induced distortions and recover task-relevant information. 
Meanwhile, the purple line with triangle markers represents the framework employing only observation domain generalization. 
Its lower classification accuracy compared with the proposed framework indicates that wireless channel impairments cannot be neglected, even when observation domain shifts are effectively mitigated. 
The modest gain from channel characterization adaptation is because the evaluation is limited to the simple AWGN channel.

\begin{table}[t] \vspace{-0.2 cm}
    \centering
    \includegraphics[width=8.8cm]{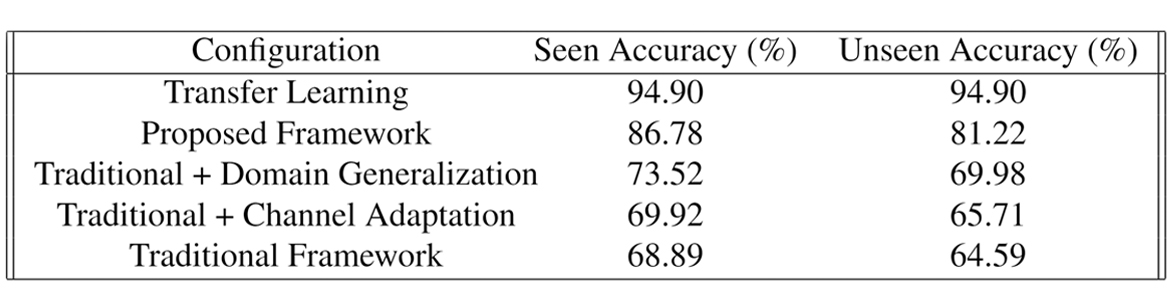}\vspace{-0 cm}
    \caption{Target-domain ablation results at an SNR of 3 dB. }  
    \label{tab:ablation}    \vspace{-0 cm}
\end{table}

To showcase the contribution of each proposed component, an ablation study at SNR of 3 dB is presented in Table \ref{tab:ablation}. 
The proposed framework achieves classification accuracies of $86.78\%$ and $81.22\%$ on the seen and unseen target domains, respectively, outperforming the traditional framework by $17.89\%$ and $16.63\%$. When channel characterization adaptation is removed, the classification accuracies decrease to $73.52\%$ and $69.98\%$, respectively. Similarly, removing observation domain generalization reduces the accuracies to $69.92\%$ and $65.71\%$. These results demonstrate that both components contribute to improving the semantic communication performance, while their joint implementation provides the additional gains in the task classification accuracy. 

\begin{figure}[t] \vspace{-0 cm}
    \centering
    \includegraphics[width=7 cm]{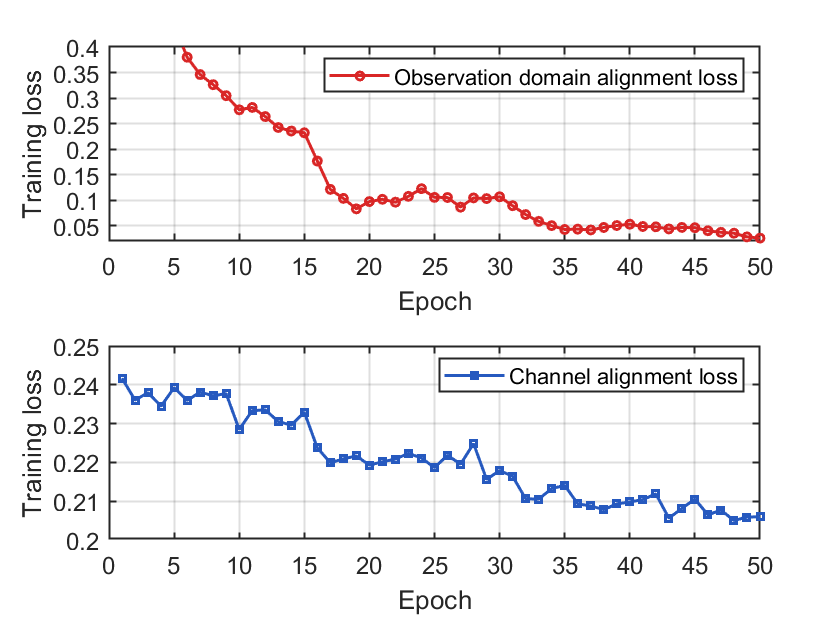}\vspace{-0.2cm}
    \caption{Training losses of observation domain alignment and channel adaptation. 
        }\vspace{-0.2 cm}
        \label{fig_loss}
\end{figure}

\begin{figure}[t] \vspace{-0.2cm}
    \centering
    \includegraphics[width=9 cm]{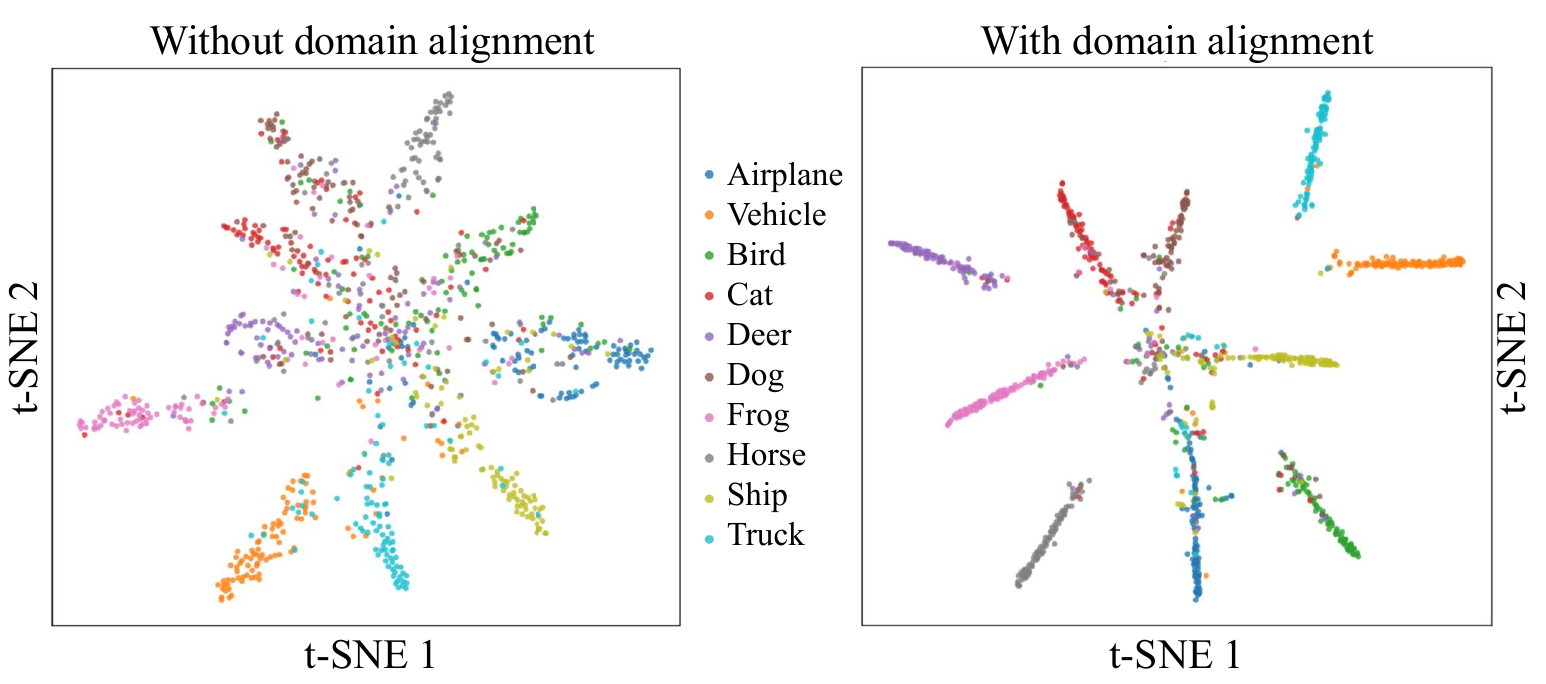}\vspace{-0.2cm}
    \caption{Task-oriented semantic representations extracted from unseen-domain inputs before and after domain alignment.} 
    \label{fig:tsne}    \vspace{-0.1cm}
\end{figure}

Fig. \ref{fig_loss}  presents the domain alignment and channel adaptation losses over $50$ training epochs.  
In the upper figure, the red line represents the observation domain alignment loss, which decreases from approximately $0.4$ to $0.03$, indicating reduced covariance discrepancy between source- and target-domain representations.  
In the lower figure, the blue line represents the training loss of the channel adaptation module, which  decreases from $0.24$ to $0.21$, demonstrating  closer agreement between the task predictions obtained from transmitted and recovered semantic representations. 

In addition to the quantitative evaluation, we employ t-distributed stochastic neighbor embedding (t-SNE)  \cite{tsne} to visualize the high-dimensional task-oriented semantic representations in the 2-dimension space.   
As shown in Fig. \ref{fig:tsne}, without domain alignment, 
the class feature extracted from unseen-domain inputs overlaps with each other, i.e., observation-domain shifts interfere with the extraction of semantic information. 
In contrast, after domain alignment, the features become more compact and separable across different classes. This result demonstrates that the proposed observation domain generalization module effectively suppresses domain-specific variations while preserving task-relevant semantic information, thereby improving the semantic representations on unseen target domains.

\section{Conclusion}\label{conclusion}

In this paper, we proposed a flexible and robust semantic communication framework to address observation domain shifts and wireless channel variations. 
Specifically, the observation domain generalization module extracted domain-invariant semantic representations, while the pilot-assisted channel characterization module mitigated channel impairments. 
Simulation results showed that the proposed framework improved classification accuracy by more than $18\%$ compared with the traditional semantic communication framework across both seen and unseen target domains.

\bibliographystyle{IEEEtran}
\bibliography{references}

\end{document}